\documentclass{article}

\usepackage{url,color,graphicx,wrapfig,comment,subcaption}
\usepackage{fullpage}

\newcommand{\ie}[1]{\emph{i.e.~}}
\newcommand{\etal}[1]{\emph{et al.~}}
\newcommand{\eg}[1]{\emph{e.g.~}}

\newcommand{\red}[1]{\textcolor{red}{#1}}

\begin{document}

\title{A survey about perceptions of mobility, \\to inform an agent-based simulator of modal choice}

\author{Carole ADAM\\Univ. Grenoble-Alpes, LIG, France\\\url{carole.adam@imag.fr} \and 
Benoit Gaudou\\Univ. Toulouse 1 Capitole, IRIT, France\\ \url{benoit.gaudou@ut-capitole.fr}}

\date{October 2024\\\red{This report is the English long version of a paper published in French at JFSMA 2024}}

\maketitle


\abstract{
In order to adapt to the issues of climate change and public health, urban policies are trying to encourage soft mobility, but the share of the car remains significant. Beyond known constraints, we study here the impact of perception biases on individual choices. We designed a multi-criteria decision model, integrating the influence of habits and biases. We then conducted an online survey, which received 650 responses. We used these to calculate realistic mobility perception values, in order to initialise the environment and the population of a modal choice simulator, implemented in Netlogo. This allows us to visualize the adaptation of the modal distribution in reaction to the evolution of urban planning, depending on whether or not we activate biases and habits in individual reasoning. 
\textbf{Demo paper}: Carole Adam, Benoit Gaudou. Un simulateur multi-agent de choix modal subjectif. JFSMA-JFMS. Nov 2024, Carghjese, Corsica
}


\section{Introduction}

The average daily distance traveled by the French population has increased considerably in recent decades, from 5 km in the 1950s to 45 km in 2011 \cite{viard2011eloge}. The number of cars has also increased (from 11.860 million in 1970 \cite{barre1997some} to 38.3 million in 2021 \cite{ccfa2019,insee}). A large proportion of home-work trips are made by car, even the shortest ones \cite{brutel2021voiture}. For example, in Toulouse, the car represents 74\% of the distances traveled by residents, and contributes 88\% of greenhouse gas emissions \cite{toulouse}.

The evolution of mobility is therefore an essential question, not only for the climate crisis but also for public health, knowing the negative impact of a sedentary lifestyle \cite{biswas2015sedentary}, road accidents, or atmospheric and sound pollution \cite{eea2016}. Indeed, each year 40,000 deaths are attributable to exposure to fine particles (PM2.5) and 7,000 deaths to nitrogen dioxide (NO2), which represents 7\% and 1\% of total annual mortality \cite{spf2021}; this report also concludes that the 2-month lockdown in spring 2020 in France made it possible to avoid 2,300 and 1,200 deaths respectively by reducing exposure to particles and nitrogen dioxide.

Recent years have seen an increased public awareness about global warming, and increased concern for ecology. Soft mobility therefore represents a well-known area of action to reduce one's carbon impact. Local and national institutions have tried to encourage them \cite{ref7-huyghe} in order to reduce traffic and pollution, via financial incentives (help to purchase a bicycle) or new infrastructure (cycle lanes). Thus during the pandemic, temporary cycle paths were created to compensate for the lower use of public transport \cite{rerat2022cycling}. But these public policies normally take longer to implement, and are not always well accepted by the car-loving population. So many of these temporary cycle paths were gradually returned to cars after the lockdowns \cite{barbarossa2020post}.

Public policies in favor of soft mobility result in increasing difficulties for motorists (low-emission zones, urban tolls, gasoline prices, pedestrian zones, parking difficulties, traffic jams, etc.) and in increased comfort and safety for cyclists thanks to dedicated facilities \cite{rerat2022cycling}. However, despite these changes, mobility is evolving very slowly. Some reasons for this inertia are well known: lack of alternatives (limited public transport, high cost of recent cars, etc.); individual constraints (transporting children or equipment); difficulty changing habits \cite{brette2014reconsidering,lanzini2017shedding}; individualism \cite{epprecht2014anticipating}; or even cognitive biases \cite{innocenti2013car,bcg2020}.

The SWITCH project\footnote{Simulating the transition of transport Infrastructures Toward smart and sustainable Cities: \url{https://www6.inrae.fr/switch}} aims at developing prospective tools to reflect on scenarios for transition of cities towards more sustainable mobility \cite{isaga-bias,isaga-habits,adam2022jeu}. In this context, we wish here to study the factors and obstacles, particularly psychological, of a transfer between modes for home-work journeys in an urban environment. To do this, we developed a modal choice model integrating the influence of perceptual biases and habits, implemented in a Netlogo simulator. We conducted an online survey aiming at identifying various cognitive biases in decisions, and calculating realistic values to initialise this simulator. 
The outline of the article is as follows: Section~\ref{sec:soa} introduces useful literature in psychology and sociology; Section~\ref{sec:model} presents our conceptual model of agent behavior; Section~\ref{sec:survey} describes our online survey and its results; Section~\ref{sec:simul} details the implementation of the Netlogo simulator and some experimental scenarios.

\section{Background about modal choice} \label{sec:soa}  

\subsection{Sociology of mobility}

Modal choice has been studied in sociology under several axes: the definition of user profiles, and the analysis of modal choice criteria.

\paragraph{User profiles.}
The Mobil'Air study \cite{chalabaevc} presents 4 car user profiles: open to all modes; attached to an individual mode, ensuring independence; constrained; and convinced, loving to drive, for all their journeys. This study provides percentages of the population in each category, but does not consider other modes of mobility. The study also notes the importance of constraints depending on the reason for travel (transporting children for example), and the strength of habits or routines.

Rocci \cite{rocci2007automobilite} proposes 6 other user profiles: passionate car drivers, adhering to the car; passionate car drivers in opposition to another mode; rational multimodal users, sticking to the car but sometimes using public transport; multimodal users in opposition to the car, but sometimes forced to take it; alternative mode users who are passionate about their mode; and alternative mode users in opposition to their mode. This classification is more detailed and considers other modes than the car, but does not provide population distribution statistics. We find several similarities with the previous survey, such as constrained or on the contrary passionate use of the car. Rocci also shows the inter-individual differences in perception of modes: for example, convinced car drivers tend to underestimate the price of the car, and overestimate that of public transport.

\paragraph{Decision factors.}
The environment imposes constraints on the modal decision, which also depends on each person's 'mobility capital' \cite{rocci2007automobilite} (owning a bicycle, having a driving license, being fit to cycle or walk, living close to a train or bus stop...). Beyond these constraints, everyone will evaluate different aspects, such as price, safety, or travel time. The choice must also minimise the mental load (number of connections, for instance). Public transport has an image associated with stress, risk of aggression and dependence \cite{kaufmann2010si}, unlike the car which conveys an image of autonomy. Finally, we retained 6 decision criteria \cite{ajce22a}: cost, time, praticity, safety, comfort, and ecology.

\paragraph{Habits.}
Habits are at the heart of mobility decisions \cite{brette2014reconsidering}: individuals tend to reproduce habitual decisions when they are in the same context. This process can save decision time, but also lead to decisions that are not adapted to changes in the environment, if they are not reconsidered. Habits can also modify perceptions (time, cost, etc.): thus a user accustomed to taking the car can overestimate travel time by public transport and under-estimate travel time by car \cite{rocci2007automobilite,betsch2001effects}.

Life cycle changes, during which all habits are disrupted (job change, moving, birth, etc.), are favorable moments for breaking old habits and creating new ones \cite{rothman2015hale}. Work has thus experimented with the distribution of free public transport tickets to newly arrived residents, to encourage them to abandon the car \cite{bamberg2003does}. The COVID-19 pandemic has also shown an unusual reset of habits and encouraged bicycle travel, at least temporarily \cite{barbarossa2020post}.

\subsection{Cognitive biases.}
Cognitive biases are heuristics used by our cognitive system to facilitate decision-making \cite{tversky1974judgment}. They enable rapid reasoning during stressful or complex situations, despite the incomplete or uncertain nature of the information necessary for rational decision-making. Although they are essential to our proper functioning, they can sometimes lead to irrational decision-making or errors. Innocenti \etal \cite{innocenti2013car} show that people tend to 'stick' to the car, even if it is more expensive than the metro or the bus, and explain this irrationality by the influence of cognitive biases. They conclude on the need for mobility policies to try to modify the perception of different modes of mobility.

Another study \cite{bcg2020} looks at the reasons why drivers are generally reluctant to switch from their personal car to new modes of mobility, even those proven to be more efficient. Although this study focuses on carpooling or free-floating bikes and scooters, their findings are interesting. They find that mobility decisions are influenced by various emotions and cognitive biases that are not taken into account by mobility operators. The \textbf{halo bias} pushes motorists to amplify the benefits of driving (autonomy...) and ignore its disadvantages (delays in traffic jams...). The \textbf{ambiguity bias} pushes them to prefer known risks to unknown risks, to maintain an illusion of control, and thus to avoid the uncontrollable risks posed by certain modes (delays or breakdowns of public transport). The \textbf{anchoring bias} implies that a negative first impression regarding a new mode of mobility will be retained, preventing future reuse. The \textbf{status quo bias} induces a preference to keep things as they are, in order to save cognitive load, which is similar to habits. The study also lists emotional and social factors which explain this attachment to the car: pride associated with owning a car, aggressiveness towards users of other modes \cite{delbosc2019dehumanization} and fear of becoming the target in the event of a modal switch; or fear of crime, particularly among women.

\section{Agent-based modal choice model}                  
\label{sec:model}                                         

Our objective is to develop a model of modal choice taking into account the role of the psychological factors described above, in particular habit and perceptual biases, in individual decisions. This model should make it possible to illustrate the impact of these factors on the effectiveness of urban development policies. To do this, the user will be able to modify the urban environment and observe how the proportion of users by mode evolves in response, depending on whether the decisions are biased or not.

\subsection{Conceptual model}

\paragraph{Environment.}
In our model, we consider 4 modes of mobility: car, bicycle, bus (urban public transport), or walking. They are evaluated according to the 6 criteria established above: ecology, comfort, price, time, practicality, and safety. The environment contains the objective values of the 4 modes with respect to the 6 criteria, representing the current urban infrastructure, and accessible to all agents. The environment is currently extremely simplified; we do not model the roads, buildings or real journeys. The focus of this model is on the influence of habits and perception biases.

\paragraph{Agents.}
The agents are endowed with an individual profile of preferences over the 6 criteria, in the form of a \textbf{vector of 6 priorities}. They also have a \textbf{filter} which biases their perception of the environment; this filter contains a multiplying factor for each criterion of each mode (so 24 factors), modifying the objective value perceived in the environment. Each agent therefore has its own \textbf{vector of subjective evaluations} of the modes (24 values), differing more or less from the objective values. From the priorities and evaluations of the criteria, the agent can then calculate the \textbf{scores} of the 4 modes and choose the one that best suits its profile. Each agent also has a home-work distance which constrains its choices (walking available below 7km and cycling below 15km) and they may or may not have \textbf{access} to the bus and the car. In addition, the agent maintains a list (sliding window) of the modes used for its last journeys and deduces a vector of \textbf{habits} which can influence its choice. The table~\ref{tab:attribs} summarises these attributes, and the following paragraphs detail these different modes of reasoning.

\begin{table}[ht]
\footnotesize
    \centering
    \footnotesize
    \begin{tabular}{|c|c|}
    \hline
        Current mode & Among: car, bike, bus, walk \\
        \hline
        Satisfaction & Score of current mode, 0-100\\
        \hline
        Distance & In km, between 0 and 200 \\
        \hline
        Access bus & Boolean \\
        Access car & Boolean \\
        \hline
        Priorities & Vector of 6 values 0-100 \\
        \hline
        Filter & Grid of 24 floats between 0.5 and 1.95\\
        \hline
        Values & Grid of 24 floats 0-100 \\
        \hline
        Subj. scores & Vector of 4 scores \\
        \hline
        Obj. scores & Vector of 4 scores \\
        \hline
        List of trips & List of the last 100 modes used \\
        \hline
        Habits & List of 4 frequencies 0-100\% \\
        \hline
    \end{tabular}
    \caption{Agent attributes}
    \label{tab:attribs}
\end{table}

\subsection{Decision process}

\paragraph{Perception filtering.} 

Each agent applies its personal perception filter on the values of the modes on the criteria. Concretely, this filter is a grid of deviation factors, associated with each mode and each criterion (there are therefore 24 factors). As the urban environment evolves, the agent perceives it but in a biased way. For each mode and each factor, it applies its deviation factor to the real value observed in the environment, to obtain its subjective perception of the value of this mode on this criterion. Each mode is associated to a filter stereotype.

This filter is dynamic, interacting with habits (see Section~\ref{sec:soa}). When using a mode for the first time, the agent has a fairly objective perception of it, or even biased by its previous mode if it had another habit. Then as it reuses and gets used to this new mode, the agent tends towards the new filter, over-evaluating this mode and under-evaluating the others. Concretely, the agents start with the filter stereotype corresponding to their usual mode. As they use other modes, their perception filter evolves via the calculation of an average between the filters of the different modes, weighted by habits. Thus an agent who was used to using the car is biased by the corresponding filter; if it uses the bicycle once, its perceptions remain biased by its main use of the car, but then its filter gradually tends towards the bicycle filter if it uses this mode again.

\paragraph{Rational choice.} The agent computes a score for each mobility mode, following a multi-criteria evaluation formula \cite{jacquier2021choice}. The rational mobility choice for the agent is the mode that receives the maximal score. Concretely, agent $i$ uses its individual priorities for each criterion $c$, denoted $prio_i(c)$, as well as the values of mode $m$ over each criterion $c$, denoted $val(m,c)$. These values depend on the current urban infrastructure (objective) and the agent's personal perception filter (subjective), and represent how well the agent believes that mode $m$ satisfies criterion $c$. The score of mode $m$ for agent $i$ is then obtained by the following formula:
$$score_i(m) = \sum_{c \in crits} val(m,c) * prio_i(c)$$

\paragraph{Habits.} Each agent maintains a list of modes used for its past journeys (sliding window). For the purposes of simplification, this list does not contain the context of the journey (destination, weather of the day, etc.) even if this would be relevant to specify these habits (\eg one may have a habit of taking the bike on sunny days but the bus when it rains). From this list, the agent calculates the past frequency of each mode. Depending on the frequency of its usual mode (the most used over this past window), the agent has a probability of choosing it again 'by habit', \ie of reusing its usual mode without further reflection \cite{isaga-habits}. If the habit is not triggered, then the agent rationally evaluates the scores of the 4 modes and chooses the best one, as described above. Therefore, the more frequently a mode is used, the more likely it is to be reused. A mode that has always been used in the past will be chosen again without any rational evaluation or comparison with other modes. If a mode has been used for half of the past journeys, the agent has a one-in-2 chance of using it again, and a one-in-2 chance of reassessing the situation before choosing.

\paragraph{Survey.} In order to calculate realistic values for the priorities and perception filters for each mobility mode, we conducted a survey, described in the following section.

\section{Mobility survey}                                 
\label{sec:survey}                                        

\subsection{Methodology}

In order to realistically initialise the population of our model, we conducted a survey via an online form. This was sent to various university mailing lists (students, laboratories, research groups, etc.) or via our personal networks \cite{conrad2024identifying}.

The questionnaire consists of three main parts. The first part concerns the profile of respondents (gender, home-work distance, number of weekly journeys) and their mobility habits (usual mode), without any identifying data. In a second part, participants are asked to evaluate their priorities for the 6 decision criteria identified above (ecology, comfort, financial accessibility, practicality, safety and speed), \ie to indicate the importance of each criterion in their choice of daily mobility. The third and final part concerns the participants' perceptions of the value of the modes of mobility considered (bicycle, car, public transport, and walking) over these 6 criteria, \ie to what extent they think that each mode meets each criterion (24 scores). All ratings (priorities and values) are given on a Likert scale from 0 to 10.

\subsection{Results}

We collected 650 responses to this survey, between March and July 2023. The numerical responses to this survey (priority and evaluation vectors) are available in open data \cite{ELLXJF_2024}. We provide here some statistics on our sample.

\paragraph{Modal distribution.}
The question about usual commuting mode allows us to deduce the following modal distribution in our sample: bicycle 31.38\% (n=204), car 20.62\% (n=134), bus 35.08\% (n=228), walk 12.92\% (n=84). In comparison, the national statistics for France \cite{perona2023deplacements} provide a very different distribution: bicycle 2\%, car 74\%, bus 16\%, walk 6\%. Our sample is therefore not representative at all of the national population, in particular because of the diffusion biases of the survey (academia), and geographical biases (diffusion from Grenoble, where the share of cycling is much greater than the national average\footnote{\url{https://www.grenoble.fr/uploads/Externe/73/215_984_Grenoble-2e-ville-cyclable-de-France-pour-les-actif.pdf} (press release) and \url{https://www.insee.fr/fr/statistiques/2557426} (INSEE statistics)}). However, the good coverage of each of the mobility modes studied will allow us to deduce interesting statistics.

\paragraph{Distances.}
The question about home-work distance allows us to deduce distance statistics by usual mode. We removed some aberrant responses (\eg walking 550km, driving 2000km, etc.), as well as zero distances and obtain the statistics summarised in the table~\ref{tab:dist} (with minimum, maximum, average, standard deviation, and median distance, as well as number of excluded values).

\begin{table}[ht]
    \centering
    \begin{tabular}{|c|c|c|c|c|c|c|}
      \hline
      Mode & min & max & avg & stdev & med & excl.\\
      \hline
      Bike & 1 & 50 & 6.43 & 6.68 & 5 & 5\\
      \hline
      Car & 2 & 190 & 21.29 & 23.1 & 15 & 3 \\
      \hline
      Bus & 1 & 95 & 11.16 & 13.59 & 5.55 & 8 \\
      \hline
      Walk & 0.2 & 9 & 1.80 & 1.44 & 1.5 & 6 \\
      \hline 
    \end{tabular}
    \caption{Distance statistics per mode}
    \label{tab:dist}
\end{table}

\paragraph{Accessibility.} 
Beyond the distance, not all modes are accessible to all users: ownership of a car or a bicycle, driving license, physical capacity, public transport coverage. We therefore calculated the number of respondents using each mode and having expressed inaccessibility not due to distance. Thus, a total of 14 cyclists, 21 motorists, and 13 bus users reported that walking was inaccessible; we note that this is the case even for very short distances of the order of 2 or 3 km. Cycling is declared inaccessible by 56 motorists, 51 bus users, and 16 pedestrians. The bus is inaccessible for 21 cyclists, 82 motorists, and 7 pedestrians. Finally, the respondents for whom a car is not available are 61 who use the bicycle, 131 who take the bus, and 42 who walk; this probably reflects a high proportion in our sample of students who do not yet have a car and/or a license. It is important to note that these inaccessibilities are purely declarative and can be subjective: a user can overestimate the time the journey would take on foot or by bike to consider it inaccessible, or find the bus too complicated to be passable. The table~\ref{tab:access} summarises the percentages of each category of users who do not have access to the bus or the car. To simplify, we consider walking and cycling as accessible unless limited by distance.

\begin{table}[ht]
    \centering
    \footnotesize
    \begin{tabular}{|c|c|c|}
    \hline
        Usual mode & No car access & No bus access \\
        \hline
        Car & 0 & 61.19\% \\
        \hline
        Bus & 57.46\% & 0 \\
        \hline
        Bike & 29.9\% & 10.29\% \\
        \hline
        Walk & 50\% & 8.33\% \\
        \hline
    \end{tabular}
    \caption{Percentage of users per mode who cannot access car / bus}
    \label{tab:access}
\end{table}

\subsection{Profiles}

\paragraph{Mean priorities.}
The table~\ref{tab:priorites} summarises the average priorities for the 6 criteria, in the total population (n=650) and by usual mode. We observe marked differences between users of different modes, for example a very low priority of motorists for ecology and price, of cyclists for safety, or of pedestrians for time.

\footnotesize
\begin{table}[ht] 
\centering 
\begin{tabular}{|c|c|c|c|c|c|} 
\hline 
 & All & Bike & Car & Bus & Walk \\ 
\hline\hline 
Ecology & 7.08 & 8.3 & 5.65 & 6.76 & 7.27\\ 
 \hline 
Comfort. & 7.1 & 7.31 & 7.19 & 6.75 & 7.35\\ 
 \hline 
Price & 6.97 & 7.08 & 5.63 & 7.44 & 7.58\\ 
 \hline 
Practicality & 8.27 & 8.54 & 8.57 & 7.81 & 8.42\\ 
 \hline 
Time & 7.47 & 7.68 & 7.79 & 7.37 & 6.7\\ 
 \hline 
Safety & 6.2 & 5.37 & 6.72 & 6.46 & 6.67\\ 
 \hline 
\end{tabular} 
\caption{Mean priorities for criteria, per usual mode of respondents} 
\label{tab:priorites} 
\end{table} 
\normalsize

\paragraph{Evaluations.}
Individuals also differ in their perception of the values of the modes on the criteria. The Table~\ref{tab:evals} indicates the average values of the 4 modes on the 6 criteria, among the overall population, then among the users of this mode, and among the non-users. We notice marked differences again, with users generally over-evaluating their mode compared to the rest of the population. Furthermore, we can notice that these deviations in evaluation are aligned with the deviations in priorities: thus cycling is evaluated as very unsafe, walking as very slow, and the car as very expensive and not very ecological. From there, we calculated the average deviation between the value of each mode on each criterion, for its users, compared to the 'objective' value taken as the median over all responses. This constitutes the perception filter prototype specific to users of this mode.

\begin{table}[ht]
\footnotesize
    \centering 
    \begin{subtable}{0.3\textwidth}
        \begin{tabular}[t]{|c|c|c|c|c|}
        \hline
        Criterion & All & U. & Non-u. \\ 
        \hline
        Ecology & 9.21 & 9.56 & 9.05 \\
            \hline
        Comfort & 6.03 & 7.39 & 5.4 \\
            \hline
        Price & 7.74 & 8.54 & 7.37 \\
            \hline
        Practicality & 6.63 & 8.23 & 5.9 \\
            \hline
        Time & 6.6 & 7.98 & 5.96 \\
            \hline
        Safety & 4.62 & 5.38 & 4.28 \\
            \hline
        \end{tabular}
        \caption{Bicycle (n=204) }
    \end{subtable}
    \begin{subtable}{0.3\textwidth}
        \begin{tabular}[t]{|c|c|c|c|c|}
        \hline
        Criterion & All & U. & Non-u. \\ 
        \hline 
        Ecology & 1.81 & 2.52 & 1.63 \\
            \hline
        Comfort & 7.69 & 8.51 & 7.47 \\
            \hline
        Price & 2.68 & 3.84 & 2.38 \\
            \hline
        Practicality & 6.32 & 8.32 & 5.81 \\
            \hline
        Time & 6.76 & 8.21 & 6.38 \\
            \hline
        Safety & 7.29 & 7.69 & 7.19 \\
            \hline
        \end{tabular}
        \caption{Car (n=134)}
    \end{subtable}
\strut \centering
    \begin{subtable}{0.3\textwidth}
        \begin{tabular}[t]{|c|c|c|c|c|}
        \hline
        Criterion & All & U. & Non-u. \\ 
        \hline 
        Ecology & 7.43 & 7.77 & 7.25 \\
        \hline
        Comfort & 5.83 & 6.46 & 5.49 \\
        \hline
        Price & 6.87 & 7.25 & 6.66 \\
        \hline
        Practicality & 5.78 & 7.2 & 5.0 \\
        \hline
        Time & 5.57 & 6.81 & 4.91 \\
        \hline 
        Safety & 7.46 & 7.37 & 7.5 \\
        \hline
    \end{tabular}
    \caption{Bus (n=228)}
    \end{subtable} 
    \begin{subtable}{0.3\textwidth}
        \begin{tabular}[t]{|c|c|c|c|c|}
        \hline
        Criterion & All & U. & Non-u. \\ 
        \hline 
        Ecology & 9.81 & 9.74 & 9.83 \\
        \hline
        Comfort & 6.7 & 8.12 & 6.49 \\
        \hline
        Price & 9.75 & 9.79 & 9.74 \\
        \hline
        Practicality & 5.99 & 8.01 & 5.69 \\
        \hline
        Time & 2.98 & 4.96 & 2.69 \\
        \hline
        Safety & 6.77 & 7.12 & 6.72 \\
        \hline
    \end{tabular}
    \caption{Walk (n=84)}
    \end{subtable}  
 
    \caption{Mean evaluation of modes over criteria}
    \label{tab:evals}
\end{table}

\paragraph{Scores of modes.}
The table~\ref{tab:stats-modes} summarises statistics about the mobility mode scores (calculated from individual responses with our multi-criteria formula)~: average over all responses, standard deviation, median among all responses, but also average among users vs. among non-users (responses whose usual mode is one of the 3 other modes). We notice again that the average score of a mode is much higher among its users than among non-users. This can be explained in several ways: on the one hand, people choosing a mode are those who evaluate it as better; and conversely, biases such as halo or \emph{a posteriori} rationalisation can lead to exaggerating the advantages and ignoring the disadvantages to confirm one's modal choice. Thus in average among the users of a mode, the chosen mode always obtains the best score among the 4, even if this is not always the case for each individual.

\begin{table}[ht]
    \centering
    \begin{tabular}{|c|c|c|c|c|c|}
        \hline
        Mode & Avg & Stdev & Med & U. & Non-u. \\
        \hline
         Bike & 6.85 & 1.66 & 7.06 & 8.11 & 6.27 \\
         \hline
         Car & 5.41 & 1.75 & 5.47 & 6.93 & 5.01 \\ 
         \hline
         Bus &  6.43 & 1.47 & 6.62 & 7.21 & 6.01 \\ 
         \hline
         Walk & 6.90 & 1.52 & 7.07 & 8.00 & 6.73 \\ 
         \hline
    \end{tabular}
    \caption{Statistics about modes scores: mean, standard deviation, median, mean score among users vs non-users}
    \label{tab:stats-modes}
\end{table}

\paragraph{Discussion.} 
The results of this survey were analysed in more detail elsewhere to identify different cognitive biases at work in modal choice. In the following we choose to focus on perception biases: over-evaluation or under-evaluation of the value of the modes on the criteria, depending on each person's habits. We therefore use the values collected via this survey and the statistical calculations presented above, to calibrate a simulator of the population's modal choices in a changing urban environment. This is described in the next section.


\section{Modal choice simulator}                  
\label{sec:simul}                                 

We implemented this conceptual model of modal choice in a Netlogo simulator \cite{netlogo99}.  


\subsection{Initialisation.}

\paragraph{Environment.}
The environment is extremely simplified. It contains neither buildings nor roads, the movements of the agents are not simulated. We are only interested in their modal decisions. Consequently, the environment only contains the numerical evaluation of the 4 modes over the 6 criteria, considered as the objective values, accessible to the agents. These values are initialised from the survey in the following way: we calculated the median evaluation of all modes on all criteria, by usual mode; we then calculated the average between these 4 different visions, weighing them by the proportion of each mode in the French population \cite{perona2023deplacements}; this allows us to adjust the average value in our non-representative sample.

\paragraph{Population.}
We used these same statistics to initialise a population of 200 agents, including 2\% cyclists, 74\% motorists, 16\% bus users, and 6\% pedestrians. Each agent is initialised with its usual mode, and attributes depending on this mode. Its list of priorities for the 6 criteria is calculated from the average priorities over users of this mode on our survey (Tableau~\ref{tab:priorites}), to which we applied an empirical random variation between minus 20 and plus 20\% in order to obtain heterogeneous profiles. Its trip list initially contains only this mode, and its perception filter is the prototype for this mode. Its home-work distance is initialised by a random draw with a Gaussian distribution parameterised by the mean and the standard deviation calculated for this mode (see Table~\ref{tab:dist}); this distance determines its access to walking and bike. Its access to the car and the bus is drawn at random according to the statistics by mode (see table~\ref{tab:access}). The number of agents created makes it possible to smooth out the effect of randomness in behaviour to observe effects at the macroscopic level.

\subsection{Interface}
\begin{figure}
\centering
    \includegraphics[scale=0.4]{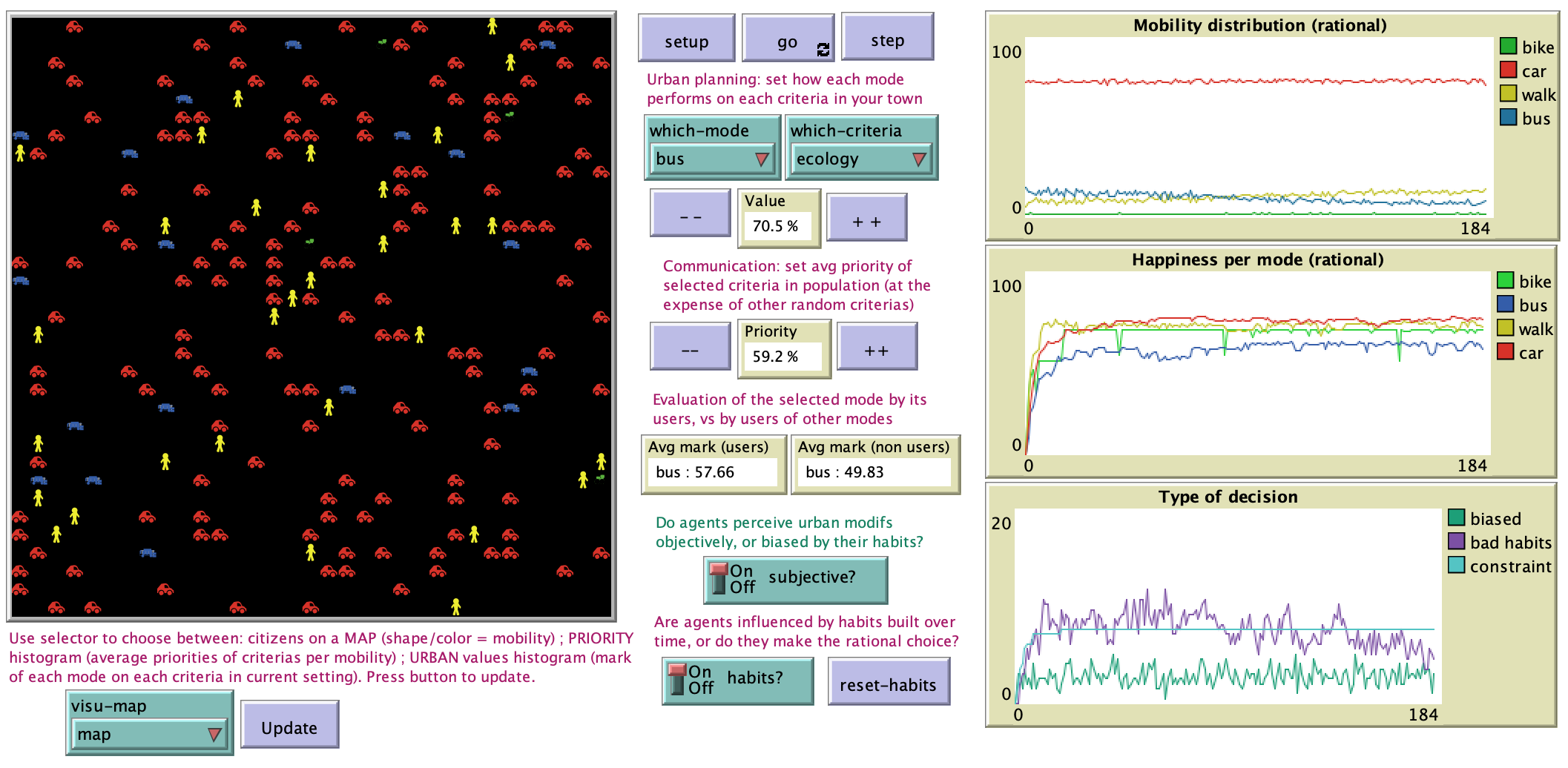}
    \caption{Netlogo simulator interface}\label{fig:screen}
\end{figure}

Figure~\ref{fig:screen} shows the Netlogo simulator interface. The user can interact by modifying the urban layout in an abstract way, by (de)activating biases and habits, and can visualise their impact on the output indicators.

\paragraph{Input parameters.}\label{parag:input}
The interface firstly allows the user to modify the urban layout, in an abstract way, that is to say by directly controlling the objective values of each mode on each criterion. Two drop-down menus allow them to select a mode and a criterion, view its current value, and decrease or increase its value. These abstract modifications have the advantage of capturing all concrete urban policies, but have the disadvantage of also allowing modifications that have no concrete meaning. For example, increasing cycling safety can correspond to creating cycle paths separated from cars; increasing the speed of the bus can correspond to a higher frequency, or to dedicated bus lanes to avoid traffic jams; reducing the financial accessibility of the car can capture the increase in the price of gasoline, or the price of tolls or parking; but reducing the ecology of walking makes no sense. To improve this simulator, it will be necessary to propose concrete actions that the player can apply, and which will then be translated into the values of these criteria. But for now, for the purpose of exploring the effect of biases, these abstract actions allow greater control of values.

Along the same lines, the user can also modify the average priority of criteria in the population, in order to simulate communication campaigns, for example to promote ecology or road safety. Once again, this is a simplification aimed at providing more control over these values to visualise their influence on the indicators at the macroscopic level. Finally, with the idea of exploring the impact of individuals' biases and habits, these two aspects of reasoning can be activated or deactivated by the user. In the absence of habits, the agents evaluate the 4 modes at each time to choose the best one. In the absence of biases, the agents use the objective unfiltered values of the modes to calculate their scores.

\paragraph{Dynamic.}
Once the population is initialized, each agent is represented with a shape and color corresponding to its current mode of mobility. At each time step, all the agents perform a random draw to find out if they activate their habit or if they re-evaluate the 4 modes, and update their attributes accordingly. In addition, we have added the possibility of a random event (probability 1\%) preventing an agent from taking its usual mode, in order to simulate possible breaks in habits (car broken down, bike punctured, weather not compatible with walking, bus strike...). The user can also reset the habits of all the agents, to simulate a global crisis (\eg pandemic): the vector of past journeys is emptied and the frequencies reset to 0, the agents are therefore forced to evaluate the available modes rationally until they build up new sufficiently strong habits. Urban layout (mode values) and priorities can be changed while the simulation runs. The left map allows three visualizations (to choose from a drop-down menu): the agents with their mode, the histograms of priorities by mode, and the histograms of values by mode, allowing the user to track the effect of their actions.

\paragraph{Output indicators.}
Several graphs allow the user to visualise the evolution of macroscopic indicators. The first plot displays the modal distribution, \ie the percentage of agents using each mode over time. The second plot presents the average satisfaction by mode, \ie the average rating given by users to their current mode. Finally, the third plot shows counts of the types of decisions taken by the agents: decisions out of habit which are contrary to the best rational choice; decisions based on biased perceptions, which differ from the decision that would be made with an objective perception; and constrained decisions, that is, when the preferred choice is not available to the agent.

\subsection{Experiments}
In this article, we focused on the qualitative exploration of the evolution of modal choices within a simulation according to parameter evolution scenarios. Indeed, the choices of agents being path dependent (due to the progressive construction of habits and the resulting biases), it seemed more important to us initially to study the evolution of choices over the course of a simulation rather than conducting an exhaustive exploration of the impact of the parameters.

\paragraph{Cycling lanes.}
In a first scenario, we want to illustrate the modal transfer when the environment evolves, and its obstacles, notably habit. To do this, we launch a simulation then gradually increase the safety of the bike (starting value 34\%, +5 points every 20 time steps). Initially the modal distribution remains stable, however the number of biased decisions increases (the perception filter prevents some agents from completely perceiving the improvement of the bicycle), as well as the number of constrained decisions (concerning the agents for whom the bicycle becomes better but is not accessible because they live too far from their work). Gradually, some bus users are converting to cycling but progress is very slow. If we reset habits during this evolution, we observe an immediate transfer from the 3 other modes (especially the car and walking) to cycling. In fact, users deprived of their habits are forced to re-evaluate modes and can therefore realise whether cycling has become better than their usual mode. The number of decisions per habit drops to 0, and the number of biased decisions increases, concerning agents whose perception filter still prevents them from considering the bicycle as better. The number of biased decisions then falls when these agents rebuild a habit that takes precedence over the biased evaluation.

\paragraph{Car comfort reduction.}
In a second scenario, we want to show the impact of constraints and habits on the modal shift. We launch the simulation without perceptual bias but with only the habits, to isolate their effect. We then gradually reduce the comfort of the car (initial value 86\%, -5 points per 20 time steps), to simulate the increase in difficulties (traffic jams, less parking spots, etc.). We observe a gradual shift of motorists with the shortest home-to-work distances towards walking. We notice that the number of forced choices increases, because motorists living further away are forced to continue using the car while its score decreases. If comfort is further reduced, motorists living further away then gradually switch to the bus, less well rated than walking but the only option. If we reset the habits, then the shift happens instantly. Ultimately, only constrained motorists remain, who continue to use the car because they do not have access to the bus.

\paragraph{Biased perception of bus.} 
In this scenario we want to show the impact of the perception filter, by launching the same simulation with then without perception bias. With filter enabled and in the initial layout, the proportions of users of each mode remain stable, corresponding to the results of the survey. On the other hand, by deactivating the filter, we observe that the proportion of bus users decreases gradually (or even instantly if we also deactivate habits). Indeed, the initial layout is not very favorable to the bus (the plot shows that its users have the lowest satisfaction of the 4 modes), often chosen by default because of distance constraints or access to other modes. The perception bias therefore makes it possible to rationalise this choice \emph{a posteriori} to improve the satisfaction felt.


\section{Conclusion} \label{sec:cci}   

In this article, we presented a modal choice model, based on a multi-criteria evaluation, with individual priorities and biased evaluations, and integrating habits. We have described the results of a survey to calculate the parameter values of this model, these responses being published as open data \cite{ELLXJF_2024}. We finally introduced a Netlogo modal choice simulator implementing this model, and presented some use case scenarios. This simulator is extremely simplified but allows the visualisation of the impact of perception biases and habits, and therefore shows the importance of considering them in the development of urban development policies. Future work will enrich the simulator with other decision factors, such as social pressure.


\paragraph{Acknowledgements.} 
The survey was carried out as part of Chloé Conrad's M1 internship \cite{conrad2024identifying}. This work is funded by the ANR (French National Research Agency) as part of the SwITCh project, under number ANR-19-CE22-0003.

\end{document}